\documentclass[a4paper]{article}

\date{2. JUN 2013}
\usepackage{icrc2013}
\newcommand{\etal}{\MakeLowercase{\textit{et al. }}} 
\title{Vectorial Radio Interferometry with LOPES 3D}
\shorttitle{D. Huber \etal LOPES 3D}
\authors{D.~Huber$^{1}$,  W.D.~Apel$^{2}$, J.C.~Arteaga-Vel\'azquez$^{3}$, L.~B\"ahren$^{4}$, K.~Bekk$^{2}$, M.~Bertaina$^{5}$, P.L.~Biermann$^{6,2}$, J.~Bl\"umer$^{1,2}$, H.~Bozdog$^{2}$, I.M.~Brancus$^{7}$, E.~Cantoni$^{5,8}$, A.~Chiavassa$^{5}$, K.~Daumiller$^{2}$, V.~de~Souza$^{9}$, F.~Di~Pierro$^{5}$, P.~Doll$^{2}$, R.~Engel$^{2}$, H.~Falcke$^{4,10,6}$, B.~Fuchs$^{1}$, D.~Fuhrmann$^{11}$, H.~Gemmeke$^{12}$, C.~Grupen$^{13}$, A.~Haungs$^{2}$, D.~Heck$^{2}$, J.R.~H\"orandel$^{4}$, A.~Horneffer$^{6}$, T.~Huege$^{2}$, P.G.~Isar$^{14}$, K-H.~Kampert$^{11}$, D.~Kang$^{1}$, O.~Kr\"omer$^{12}$, J.~Kuijpers$^{4}$, K.~Link$^{1}$, P.~{\L}uczak$^{15}$, M.~Ludwig$^{1}$, H.J.~Mathes$^{2}$, M.~Melissas$^{1}$, C.~Morello$^{8}$, J.~Oehlschl\"ager$^{2}$, N.~Palmieri$^{1}$, T.~Pierog$^{2}$, J.~Rautenberg$^{11}$, H.~Rebel$^{2}$, M.~Roth$^{2}$, C.~R\"uhle$^{12}$, A.~Saftoiu$^{7}$, H.~Schieler$^{2}$, A.~Schmidt$^{12}$, F.G.~Schr\"oder$^{2}$, O.~Sima$^{16}$, G.~Toma$^{7}$, G.C.~Trinchero$^{8}$, A.~Weindl$^{2}$, J.~Wochele$^{2}$, J.~Zabierowski$^{15}$, J.A.~Zensus$^{6}$}

\afiliations{
 $^1$Institut f\"ur Experimentelle Kernphysik, Karlsruher Institut f\"ur Technologie (KIT), Germany\\
 $^2$Institut f\"ur Kernphysik, Karlsruher Institut f\"ur Technologie (KIT), Germany\\
 $^3$Universidad Michoacana, Instituto de F\'{\i}sica y Matem\'aticas, Morelia, Mexico\\
 $^4$Department of Astrophysics, Radboud University Nijmegen, The Netherlands\\
 $^5$Dipartimento di Fisica dell' Universit\`a Torino, Italy\\
 $^6$Max-Planck-Institut f\"ur Radioastronomie Bonn, Germany\\
 $^7$National Institute of Physics and Nuclear Engineering, Bucharest, Romania\\
 $^8$Osservatorio Astrofisico di Torino, INAF Torino, Italy\\
 $^9$Universidade S$\tilde{a}$o Paulo, Instituto de F\'{\i}sica de S$\tilde{a}$o Carlos, Brasil\\
 $^{10}$ASTRON, Dwingeloo, The Netherlands\\
 $^{11}$Fachbereich Physik, Universit\"at Wuppertal, Germany\\
 $^{12}$Institut f\"ur Prozessdatenverarbeitung und Elektronik, Karlsruhe Institute of Technology (KIT), Germany\\
 $^{13}$Fachbereich Physik, Universit\"at Siegen, Germany\\ $^{14}$Institute of Space Science, Bucharest, Romania\\
 $^{15}$National Centre for Nuclear Research, Department of Cosmic Ray Physics, {\L}\'od\'z, Poland\\
 $^{16}$Department of Physics, University of Bucharest, Bucharest, Romania}

\email{daniel.huber@kit.edu}

\email{daniel.huber@kit.edu}

\abstract{One successful detection technique for high-energy cosmic rays is based on the radio signal emitted by the charged particles in an air shower \cite{Haungsradio}. The LOPES experiment \cite{FalckeNature2005} at Karlsruhe Institute of Technology, Germany, has made major contributions to the evolution of this technique. LOPES was reconfigured several times to improve and further develop the radio detection technique. In the latest setup LOPES consisted of 10 tripole antennas. With this, LOPES 3D \cite{lopes3d} was the first cosmic ray experiment measuring all three vectorial field components at once and thereby gaining the full information about the electric field vector. We present an analysis based on the data taken with special focus on the benefits of a direct measurement of the vertical polarization component. We demonstrate that by measuring all polarization components the detection and reconstruction efficiency is increased and noisy  single channel data can be reconstructed by utilising the information from the other two channels of one antenna station.}

\begin{document}
\maketitle
\section{Introduction}
Observing cosmic rays at energies $\geq10^{17}$\,eV is a challenging task since only indirect measurements of air showers can be performed. For these measurements different techniques can be used. One of these techniques is the measurement of the radio emission caused by the deflection and time variation of the number of charged particles in an air shower. This emission is not absorbed in the atmosphere and is only influenced by strong atmospheric electric fields which are present during thunderstorms \cite{lopesthunderstrom}. Since the radio pulse is generated during the complete air shower development the radio detection is very sensitive to  the energy of the primary particle \cite{nunzia}. This sensitivity combined with a very high duty cycle gives the radio technique a unique position in cosmic ray detection. Thus it is of high interest to further develop and fully exploit the potential of this technique. The LOPES radio antenna array was reconfigured to be able to measure all three components of the electric field vector from radio emission directly via digital radio interferometry. In this article we present the updated treatment of the antenna gain and the improvement of the interferometrical method when calculating the cross-correlation vectorially.

\section{LOPES 3D Setup}
LOPES 3D is the last setup of the LOPES experiment. As all configurations before it measures the radio emission from cosmic ray air showers at energies larger than $10^{17}$\,eV. It has a bandwidth of $40$ to $80$\,MHz, is situated within the KASCADE \cite{kascade} array at Karlsruhe Institute of Technology and is triggered by KASCADE and \mbox{KASCADE-Grande} \cite{kascade-grande}. LOPES 3D was stopped in January of 2013. At that time roughly $2.5$\,years of data had been taken. The outstanding feature of LOPES 3D is the deployed antenna type. At LOPES 3D for the first time an antenna that is sensitive to all components of the electric field vector is used in cosmic ray physics. This antenna consists of three crossed dipoles that are perpendicular to each other and is called a tripole.

\begin{figure}[!t]
\begin{center}
\includegraphics[width= .475\textwidth ,angle=0]{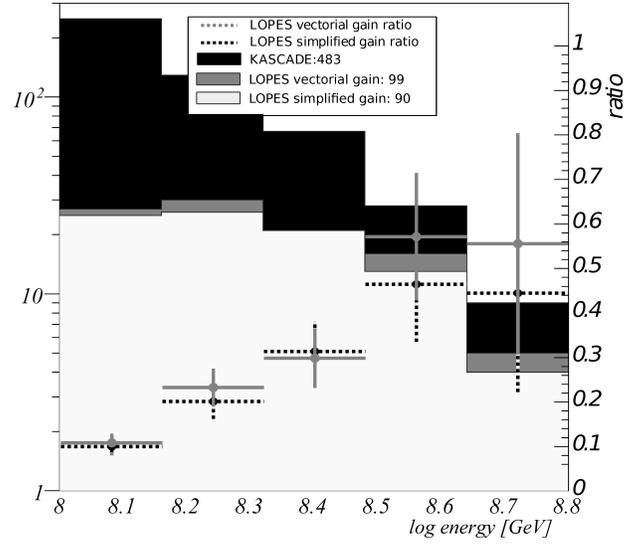}
\end{center}
\caption{Histogram of KASCADE events (black) and reconstructed LOPES events in at least one E-field component (bright grey/dark grey) and the ratio of the reconstructed events (solid/dashed points) for the absolute gain approximation according to equation \ref{equ:simplegainew} and for the vectorial gain treatment. Additional cut is applied on the geomagnetic angle $\alpha \geq 20^{\circ}$.}
\label{fig:cc_classic_ratio}
\end{figure} 

\begin{figure}[!t]
\begin{center}
\includegraphics[width= .475\textwidth ,angle=0]{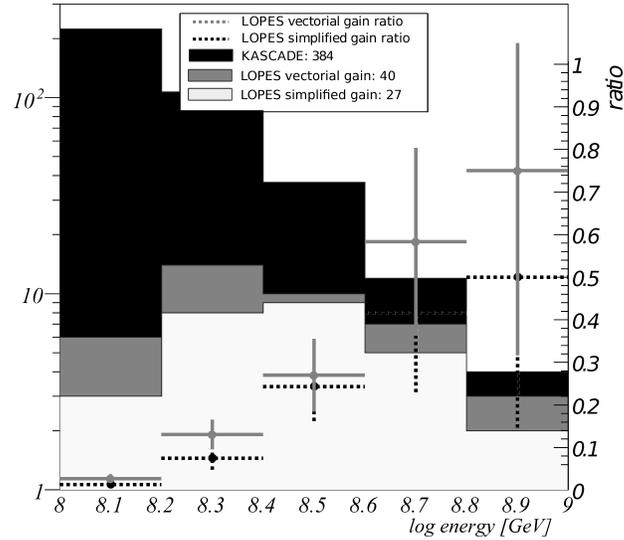} 
\end{center}
\caption{Ratio of reconstructed events for the absolute gain approximation according to equation \ref{equ:simplegainew} (black points) and for the vectorial gain treatment (grey points). Additional cut is applied on the geomagnetic angle $\alpha \geq 45^{\circ}$.}
\label{fig:lopes30vecgain}
\end{figure}

\section{Antenna gain treatment}
\label{chap:vectorialgaintreatment}
When analysing the radio emission from extensive air showers with high precision it is indispensable to understand every single component of the data acquisition in detail. One very crucial point is the gain pattern of the radio antenna used. The gain pattern describes the sensitivity of the antenna and mainly depends on: the direction from which the signal arrives at the antenna, the frequency of the signal and the orientation of the E-field vector with respect to the antenna.
The gain pattern can vary with changing ground or weather conditions. Metal parts near the antenna have an influence as well. Applying the correct gain and reconstructing the initial electric field vector \cite{radiooffline} will be explained in detail in the following. 

\subsection{Simplified antenna gain treatment}
 When measuring with just one antenna per station one way to calculate the according electric field vector component is to use a simplification. This assumes that the measured signal is completely polarized in the according direction, which is in this case east-west, see equation \ref{equ:simplegain}.

\begin{equation}
S_{\mbox{ant}} = |\vec{E}|_{\mbox{ew}}\cdot | \vec{G}(f,\theta,\phi)|
\label{equ:simplegain}
\end{equation} 

 with S being the measured voltage at the antenna foot point, E the electric field vector, and G the gain of the antenna. This can be used to approximate the east-west component $|\vec{E}|_{\mbox{ew}}$ according to equation \ref{equ:simplegainew}.

\begin{equation}
|\vec{E}|_{\mbox{ew}}= \frac{S_{\mbox{ant}}}{| \vec{G}(f,\theta,\phi)|} 
\label{equ:simplegainew}
\end{equation}

\subsection{Vectorial gain treatment}
With the measurement the E-field vector is translated from a complex vector to a scalar voltage at the antenna foot point. This implies that to reconstruct the complete vector three scalars need to be measured. Since the radio emission from cosmic ray induced air showers is a transversal electro-magnetic wave, the E-field vector in the plane perpendicular to the Poynting vector reduces to two components. Thus, if the arrival direction of the emission is known, the electric field vector can be completely calculated when measuring only two signals.
From the two measured signals:
\begin{equation}
S_{\mbox{ant1}} = \vec{E}\cdot\vec{G}(f,\theta,\phi)_{\mbox{ant1}} 
\end{equation}
\begin{equation}
S_{\mbox{ant2}} = \vec{E}\cdot\vec{G}(f,\theta,\phi)_{\mbox{ant2}}
\end{equation}

the two components of the E-field vector in the shower plane can be calculated:
 
  \begin{equation}
 E_{\mbox{ze}}=\frac{S_{\mbox{ant2}}\cdot G{^{\mbox{az}}_{\mbox{ant1}}} - S_{\mbox{ant1}}\cdot G{^{\mbox{\mbox{az}}}_{\mbox{ant2}}}}{G{^{\mbox{az}}_{\mbox{ant1}}}\cdot G{^{\mbox{ze}}_{\mbox{ant2}}} - G{^{\mbox{az}}_{\mbox{ant2}}}\cdot G{^{\mbox{ze}}_{\mbox{ant1}}}}
 \label{equ:ccvec1}
   \end{equation}
   
  \begin{equation}
 E_{\mbox{az}}=\frac{S_{\mbox{ant1}}\cdot G{^{\mbox{ze}}_{\mbox{ant2}}} - S_{\mbox{ant2}}\cdot G{^{\mbox{\mbox{ze}}}_{\mbox{ant1}}}}{G{^{\mbox{az}}_{\mbox{ant1}}}\cdot G{^{\mbox{ze}}_{\mbox{ant2}}} - G{^{\mbox{az}}_{\mbox{ant2}}}\cdot G{^{\mbox{ze}}_{\mbox{ant1}}}}
 \label{equ:ccvec2}
   \end{equation}

This completely reconstructed electric field vector then can be rotated to a fixed coordinate system, to be comparable with other measurements or simulations. For \mbox{LOPES 3D} there are $3$ channels available at each antenna station. With this \mbox{LOPES 3D} measures redundant since with a measurement with $2$ channels the E-field vector is determined. Thus either noisy or broken channels can be compensated or the reconstruction of the E-field vector can be performed $3$ times using every combination of $2$ channels. Thus if no channel is taken out of the analysis 3 independent reconstructions are available that are combined using different weights. 

\begin{figure}[!t]
\begin{center}
\includegraphics[width= .415\textwidth ,angle=0]{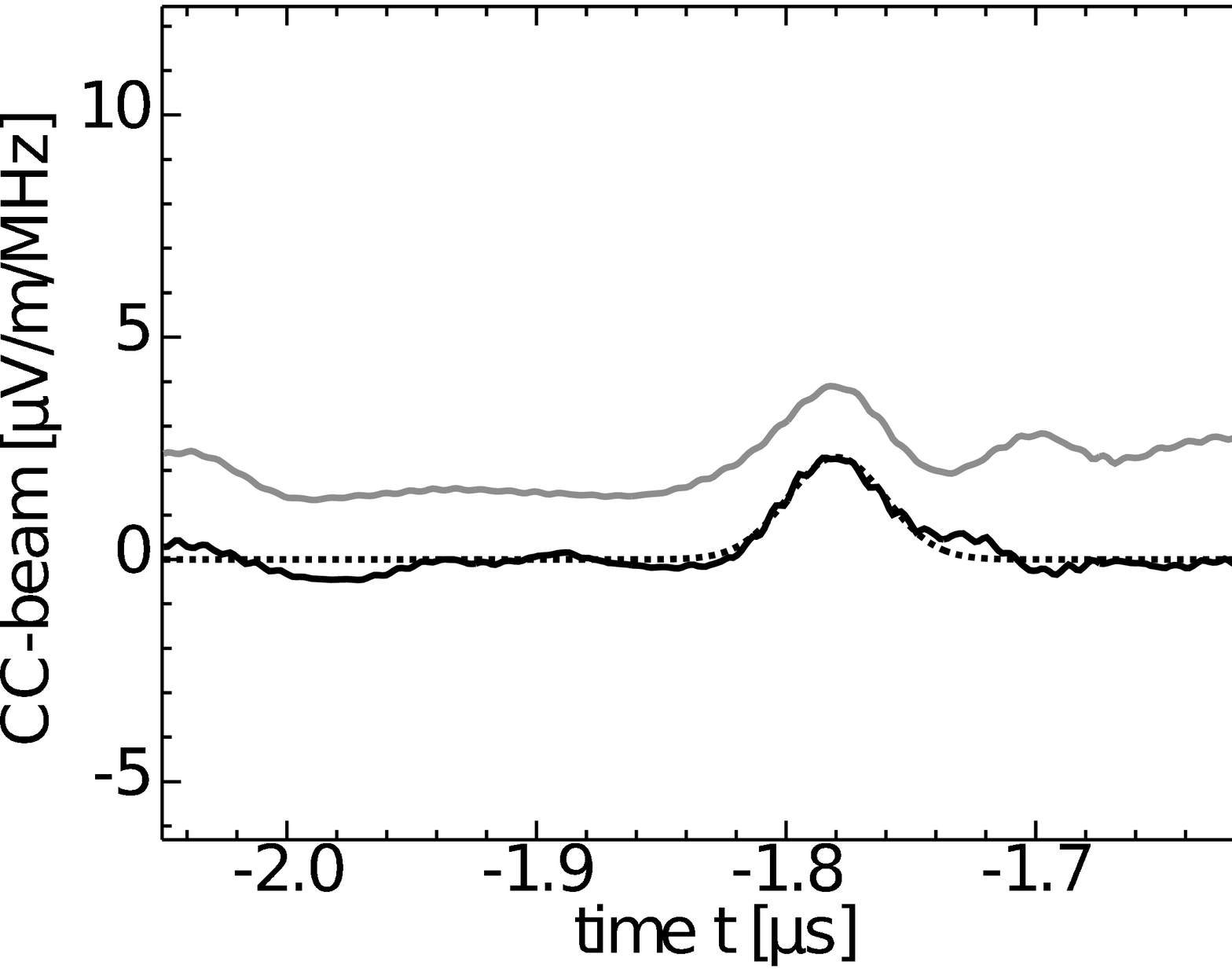} 
\put	(-160,120){\huge EW}
\\ \quad \hfill \\
\includegraphics[width= .415\textwidth ,angle=0]{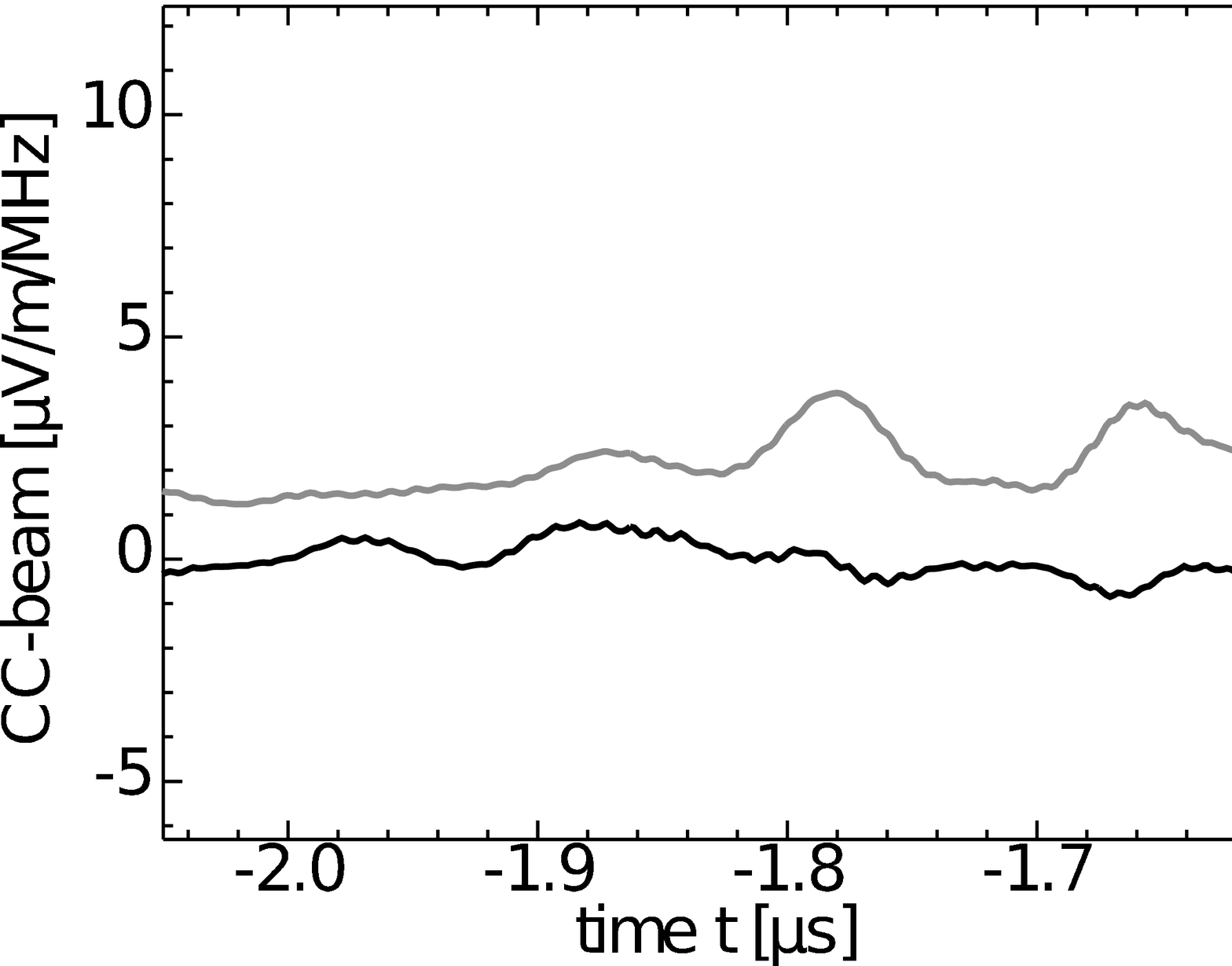}
\put	(-160,120){\huge NS}
\\  \quad \hfill \\
\includegraphics[width= .415\textwidth ,angle=0]{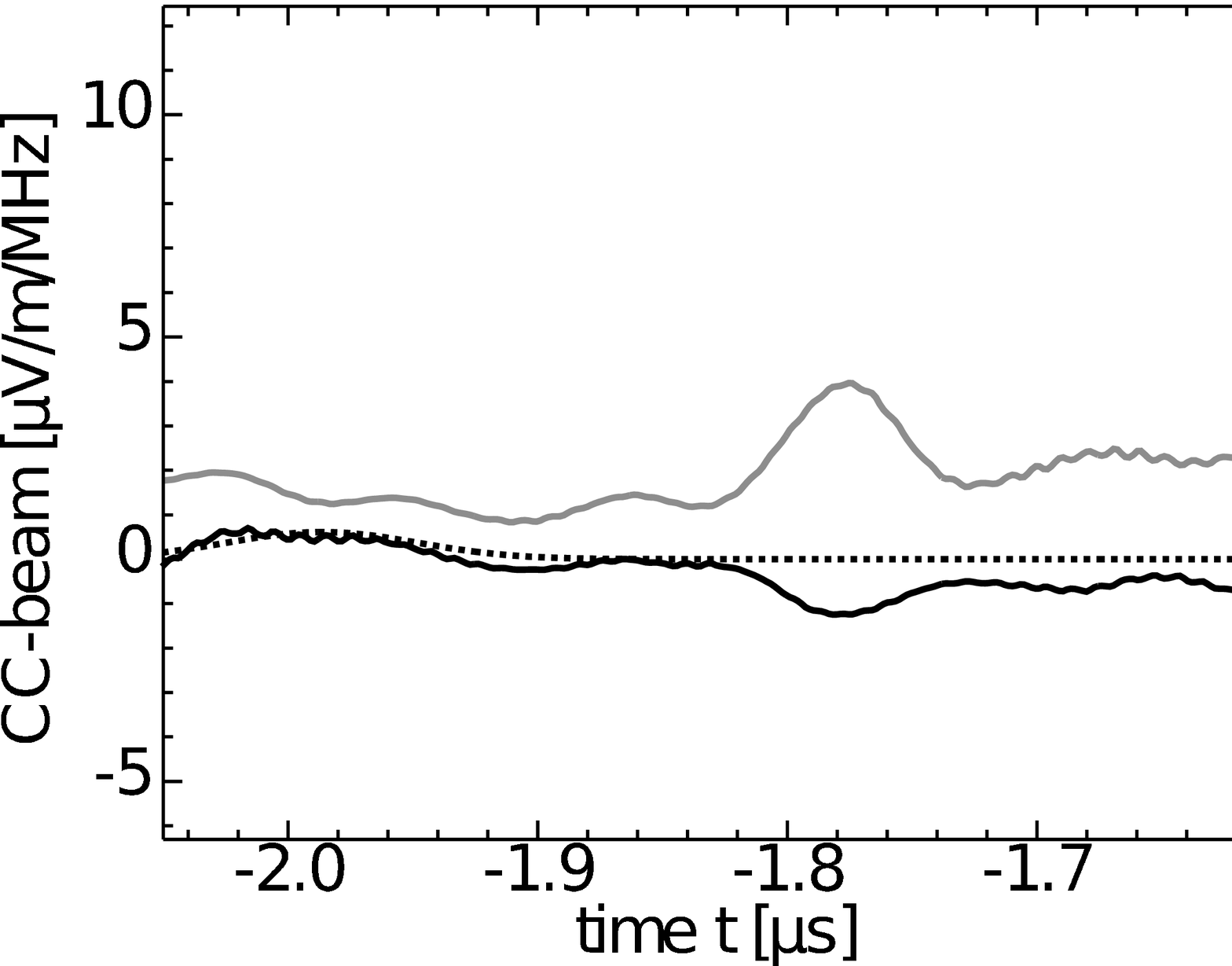}
\put	(-160,120){\huge VE}
\end{center}
\caption{CC-beam calculated with the vectorial gain, but for each component separately. The cc-beam is shown as solid black line, the power-beam as grey line and the Gaussian fit, if converged, is shown as dashed black line.}
\label{fig:ccbeamnovec}
\end{figure} 

\begin{figure}[!h]
\begin{center}
\includegraphics[width= .415\textwidth ,angle=0]{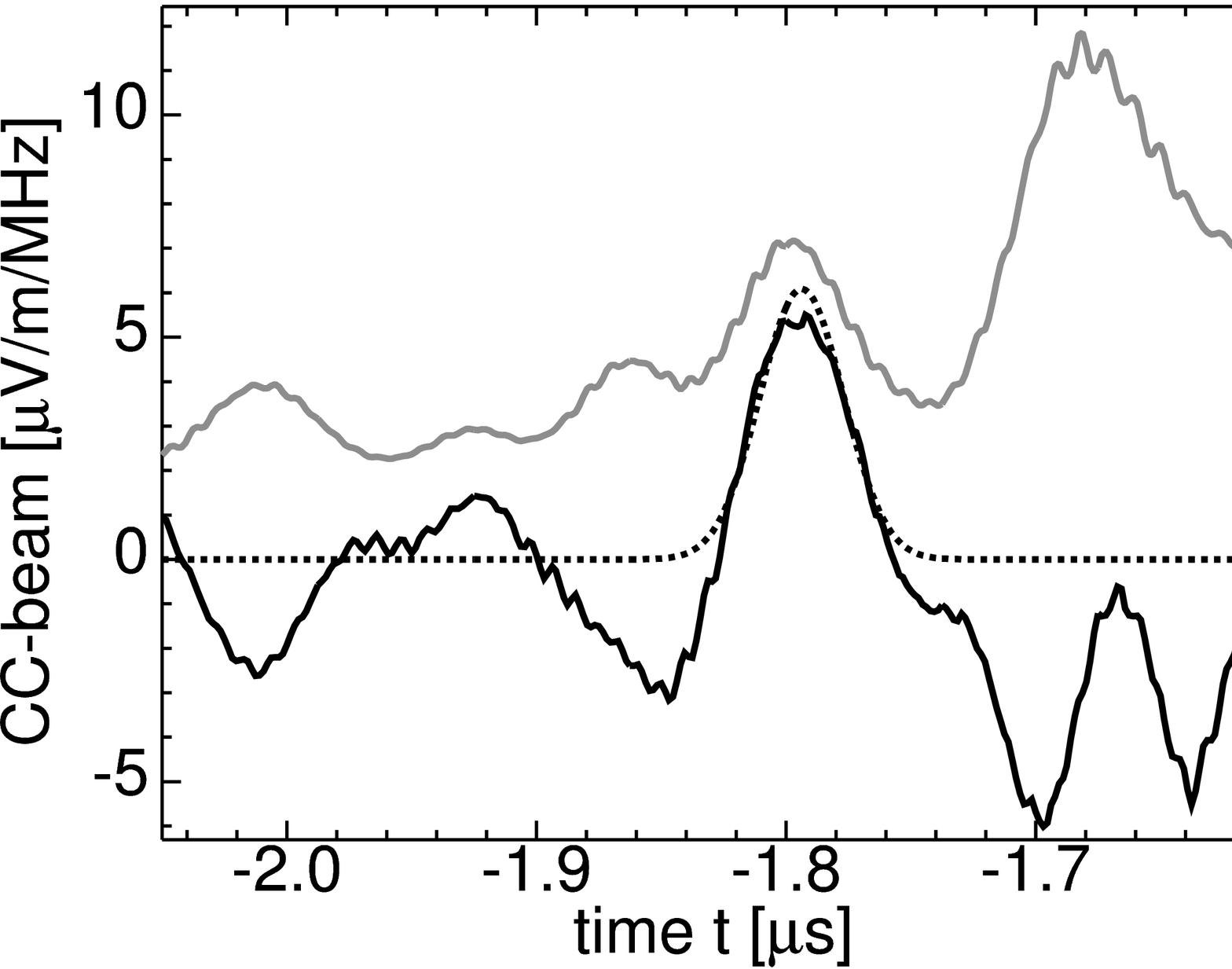}
\put	(-160,120){vectorial}
\put	(-160,110){beamforming}
\end{center}
\caption{CC-beam calculated vectorially combining information from all three orientations of the E-field with the vectorial gain treatment including phases. The cc-beam is shown as solid black line, the power-beam as grey line and the Gaussian fit as dashed black line.}
\label{fig:ccbeamvec}
\end{figure}

\section{Applying the gain vectorially to LOPES data}
In order to study the influence of the newly developed gain treatment two analyses were done. First the gain is treated according to the simplification described in equations \ref{equ:simplegainew} and \ref{equ:simplegain}, and then the cross correlation beam (cc-beam) height is checked and the event is considered as successfully reconstructed if the signal-to-noise ratio of the cc-beam is above a chosen threshold. The ratio of reconstructed events from KASCADE and LOPES using this condition is shown in figure \ref{fig:cc_classic_ratio}. The same is repeated for the gain treatment according to equations \ref{equ:ccvec1} and \ref{equ:ccvec2}. An  improvement of the reconstruction efficiency using the new gain treatment can be observed, see figure \ref{fig:cc_classic_ratio}. To have a fair comparison the reconstruction was performed for each component separately in both cases. This ensures that only the new gain treatment affects the results presented in figure \ref{fig:cc_classic_ratio}.\\ For the $5$ double polarized antenna stations of the LOPES 30 pol setup this analysis was also done, but with a higher cut on the geomagnetic angle ($\alpha \geq 45^{\circ}$) since fewer antenna stations are available. Also in this dataset a clear improvement could be observed, see figure \ref{fig:lopes30vecgain}. The higher ratio  compared to the LOPES 3D dataset originates from the stricter quality cut which increases the purity in this sample.

\section{Beamforming}
When calculating the cross correlation beam LOPES is used as a digital radio interferometer \cite{HornefferThesis2006}. The beamforming significantly increases the signal-to-noise ratio of the measured radio pulse and is the first step of any analysis performed with LOPES. Without this advanced technique it is not possible to detect the radio emission from cosmic ray induced air showers in a noisy environment like the LOPES site.

During the beamforming the time traces of the single antennas are shifted with respect to each other to gain the highest sensitivity in the source direction. After the shifting the cross-correlation of the traces is calculated to form the cc-beam or the square of the traces can be added up to form the power beam (p-beam). So far this was done for the east-west aligned antennas, the north-south aligned antennas and the vertical antennas separately, cf. equation \ref{equ:ccbeamscalar}.

\begin{equation}
\sqrt{ \left| \frac{1}{N} \sum_{i=1}^{N-1}{\sum_{i>j}^{N}{f_{i}[t]\cdot
f_{j}[t]}} \right|} 
\label{equ:ccbeamscalar}
\end{equation}
$f[t]$ denotes the time trace of one channel and N the number of channels. \\ Forming the cc-beam helps identifying even signals at the noise level. The cc-beamforming works better the more traces go in the calculation, thus on the one hand it is desirable to have as many traces as possible going in the cc-beam calculation. On the other hand, if a trace has no information on the signal and only contains noise, it will significantly harm the height of the cc-beam and in the worst case make the reconstruction of the air shower impossible. 
\\

\subsection{Vectorial beamforming with LOPES 3D}
The cc-beam calculation has been updated to work on the complete E-field vector. This is the first time a vectorial beamforming has been applied to identify air shower radio emission with LOPES. Working on the full vector uses all the information available for the reconstruction. \\ In figure \ref{fig:ccbeamnovec} the cc-beam for a LOPES 3D recorded event is shown. The beam was calculated with the vectorial gain treatment, but reconstructed for each orientation separately. The E-field vector was reconstructed using only the signals from the east-west and north-south aligned antennas. A clear coherent signal is only visible in the east-west component. The cc-beam shown in figure \ref{fig:ccbeamvec} is calculated vectorially combining information from all three orientations of the E-field and superposing all three reconstructions of the E-field vector with equal weights. The signal is clearly increased whereas the noise level stays the same. The received power gets higher since the radio noise emitted by the KASCADE photomultipliers is more prominent in the vertical oriented antennas. This example demonstrates that with vectorial beamforming the reconstruction can be improved, as long as every polarization at least contains some information of the signal. Doing a vectorial  beamforming on data which could not be reconstructed before led to $4$ additional events that are above threshold.

\subsection{Near horizontal showers with LOPES 3D}
\begin{figure}[!t]
\begin{center}
\includegraphics[width= .475\textwidth ,angle=0]{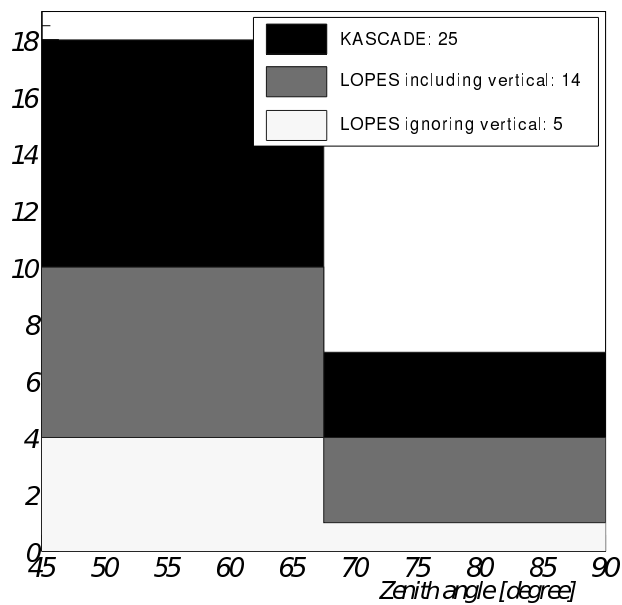} 
\caption{Events passing a chosen cut on the SNR of the cc-beam over zenith angle for a reconstruction ignoring the measurement from the vertically oriented antennas (bar filled histogram), and including these measurements (bright grey) on top of the angular distribution of the triggers.}
\end{center}
\label{fig:flatwithv}
\end{figure}

The main advantage of the LOPES 3D setup is the direct measurement with vertically oriented antennas. With this LOPES 3D is expected to be well-suited for the detection of inclined showers since the signal in the vertically oriented antennas increases. For a small subset of KASCADE-Grande triggered flat events (cut on zenith angle $\geq 45^{\circ}$) the influence of the direct measurement with vertically oriented antennas was checked. In figure \ref{fig:flatwithv}. As expected more events can be reconstructed. KASCADE-Grande is due to the flat scintilators loosing sensitivity to near horizontal showers. The high ratio of reconstructed events most probably originates from this detection bias since not all showers are triggered uniformly distributed.

\section{Conclusions}
Radio detection of cosmic rays has evolved considerably in the past years. Within this development the precision increased which made it more and more important to reduce experimental uncertainties. One crucial point is the gain pattern of the antenna. When reconstructing the E-field the gain pattern has to be taken into account vectorially. In the current analysis of the data taken with LOPES the gain and the beamforming were updated to be performed vectorially which led to a significant increase in the detection efficiency and the quality of air shower reconstruction. Including the measurement from the vertically oriented antennas in the analysis significantly increases the detection efficiency and reconstruction efficiency for inclined showers. The improvement of the vectorial beam forming based on an advanced analysis with dynamical weight calculation of the independent reconstructed E-field vector will be shown in a following analysis.

\vspace*{0.5cm} \footnotesize{
{\bf Acknowledgement:} Part of this research has been supported by grant number VH-NG-413 of the 
Helmholtz Association.}


\begin{thebibliography}{}

\bibitem
{Haungsradio}
Haungs, A., NIM A {\bf 604} S236-S243 (2009).
 
\bibitem
{FalckeNature2005}
H. Falcke et al. (LOPES Collaboration), Nature {\bf 435} (2005) 313-316.
 
\bibitem{lopes3d}
W.D. Apel et al. (LOPES Collaboration) NIM A {\bf 696} (2012) 100-109 

\bibitem{lopesthunderstrom}
W.D. Apel et al., Advances in Space Research  {\bf 48} (2011) 1295-1303 

\bibitem{nunzia}
N.~Palmieri, \etal (LOPES Collaboration), 
paper 0439, these proceedings.


\bibitem
{kascade}
T.~Antoni et al. (KASCADE Collaboration), NIM A {\bf 513} (2003) 429.


\bibitem
{kascade-grande}
W.-D.~Apel et al. (KASCADE-Grande Collaboration), NIM A {\bf 620} (2010) 490-510.


\bibitem{radiooffline}
P. Abreu et al (Pierre Auger Collaboration), NIM A {\bf 635} (2011) 92-102.


\bibitem{rothammel}
Rothammels Antennenbuch (2012) \\
ISBN 978-3-88692-033-4


\bibitem{HornefferThesis2006}
Horneffer, A., Rheinische Friedrich-Wilhelms-Universit\"at Bonn PhD Thesis (2006).




\end{thebibliography}
\end{document}